\documentclass[twocolumn,tighten,times]{aastex631}

\usepackage{color}
\usepackage{url}
\usepackage{hyperref}
\usepackage{xspace}
\usepackage{soul}

\shorttitle{X-rays from NGC 2366}
\shortauthors{Kaaret \& Prestwich}

\begin{document}

\title{X-rays from the Nearby Compact Emission Line Galaxy NGC 2366}

\author[0000-0002-3638-0637]{Philip Kaaret}
\affiliation{NASA Marshall Space Flight Center, Huntsville, AL 35812, USA}

\author[0000-0003-3484-0326]{Andrea Prestwich}
\affiliation{Center for Astrophysics, Harvard-Smithsonian, 60 Garden Street, Cambridge, MA 02138, USA}


\begin{abstract}

Luminous, compact emission-line galaxies (LCGs) are the most abundant class of confirmed Lyman continuum (LyC) emitters. An optical integral field study of the nearby LCG NGC~2366 reveals an outflow originating at the star cluster `knot B' thought to clear a channel via mechanical feedback that enables LyC escape. We observed NGC~2366 with the Chandra X-ray Observatory and detect X-ray emission from a point source coincident with the apex of the outflow at knot B. The point-like nature and variability of the X-ray emission suggests accretion onto a compact object. The accretion could produce sufficient kinetic energy to power the outflow.

\end{abstract}


\section{Introduction}
\label{sec:intro}

Understanding the mechanisms that enable Lyman continuum (LyC) radiation to escape galaxies is essential to understanding how the universe was reionized. LyC is absorbed by dust and neutral hydrogen, both of which are found in abundance in star-forming regions \citep{Hayes2010}. Feedback may be needed to blow neutral gas and dust away from the starburst to enable LyC escape. Stellar winds and supernovae are often considered as sources of mechanical power generated by star-forming regions \citep[e.g.,][]{Kimm2014}. Outflows from accretion onto compact objects in high-mass X-ray binaries are another potential source of mechanical power from star-forming regions \citep{Prestwich2015}. The outflow power is often comparable to the radiative luminosity \citep{Gallo2005}. In some systems, the mechanical power dominates the radiative luminosity by several orders of magnitude, e.g.\ the mechanical power of the microquasar in NGC~7793 is a few $10^{40}$~erg/s while the X-ray luminosity of the binary is only $7 \times 10^{36}$~erg/s \citep{Pakull2010}.

The few nearby galaxies with LyC escape serve as local laboratories that can be resolved with current X-ray and optical instrumentation. The three nearby ($z < 0.1$) galaxies that have the strongest evidence for LyC emission are compact galaxies with high star formation rates and all three host luminous, variable, and hard-spectrum X-ray sources: Haro 11 \citep{Prestwich2015,Gross2021}, Tololo 1247-232 \citep{Kaaret2017}, and Tololo 0440-381 \citep{Kaaret2022}. These X-ray sources are likely powered by accretion on to compact objects and likely produce powerful outflows that could help enable LyC escape. Study of a larger set of candidate LyC emitters shows that the presence of X-ray sources correlates
with reduced dust obscuration, again suggesting a potential link between the presence of accretion-powered objects and LyC escape \citep{Bluem2019}.

Luminous, compact, emission-line galaxies (LCGs) have high star-formation rates and are good local cognates to the compact, low-metallicity, high-star-formation rate galaxies thought to have reionized the early universe. Such galaxies with redshifts near 0.2 are referred to as `Green Peas' \citep{Cardamone2009}. However, galaxies with similar properties can be found over a wide redshift range \citep{Izotov2011ApJ728}. Remarkably, 11 out of the 11 LCGs observed with the Cosmic Origins Spectrograph (COS) on the Hubble Space Telescope (HST) in the appropriate redshift range have been confirmed as Lyman continuum emitters (LyC) by direct observation \citep[and references therein]{Micheva2019}. This makes LCGs/Green Pea analogs the most abundant class of LyC emitting galaxies.

NGC~2366 is the closest Green Pea analog, at a distance of 3.44~Mpc \citep{Tolstoy1995}, and is a candidate LyC emitter. The core star-forming region of NGC~2366 is also known as Mrk~71 or NGC~2363 and contains two star clusters known as knot A and B \citep{Drissen2000}. \citet{Micheva2019} performed optical integral-field spectroscopy of Mrk~71 and the results indicate that outflows are an important ingredient for LyC leakage. Optical line diagnostics from the spatially-resolved spectra enabled them to measure the electron temperature and density and velocity dispersion at each location in order to create maps of the sound speed, thermal broadening, and Mach number within the galaxy. The resulting map of the velocity dispersion reveals a cone structure with an apex near knot~B. The opening angle of the cone is consistent with the Mach angle calculated from the Mach number measured at knot~B. The cone falls in a region of low dust extinction. This supports a scenario in which a mechanical outflow from knot~B creates a low density channel in the interstellar medium within the galaxy enabling LyC escape.

XMM-Newton detected an X-ray source near the star-forming knots in NGC~2366 \citep{Thuan2014}. We conducted observations with the Chandra X-ray Observatory in order to resolve the X-ray emission and examine the relation of the X-ray emission to the measured outflows and stellar clusters within the galaxy. We report on the observations and our analysis in section~\ref{sec:obs} and describe the results in section~\ref{sec:results}. We discuss the results and their implications for our understanding of LyC esscape in section \ref{sec:discussion}.

\begin{table}
\caption{Chandra observations of NGC 2366}
\begin{tabular}{ccc}
\hline
  ObsID & Exposure (ks) & Start Date/Time (UTC) \\ \hline
  25228 & 30.44 & 2021-11-24 20:01:45 \\
  25737 & 29.68 & 2021-12-04 18:28:56 \\
  25738 & 26.60 & 2021-12-05 05:38:09 \\
  25736 & 17.73 & 2023-03-27 12:48:13 \\
  27257 & 12.79 & 2023-09-13 05:33:22 \\ \hline
\end{tabular} 
\label{tab:obs}
\end{table}

\begin{figure*}[tb]
\begin{minipage}[c]{0.48\textwidth} 
\centerline{\includegraphics[width=\textwidth]{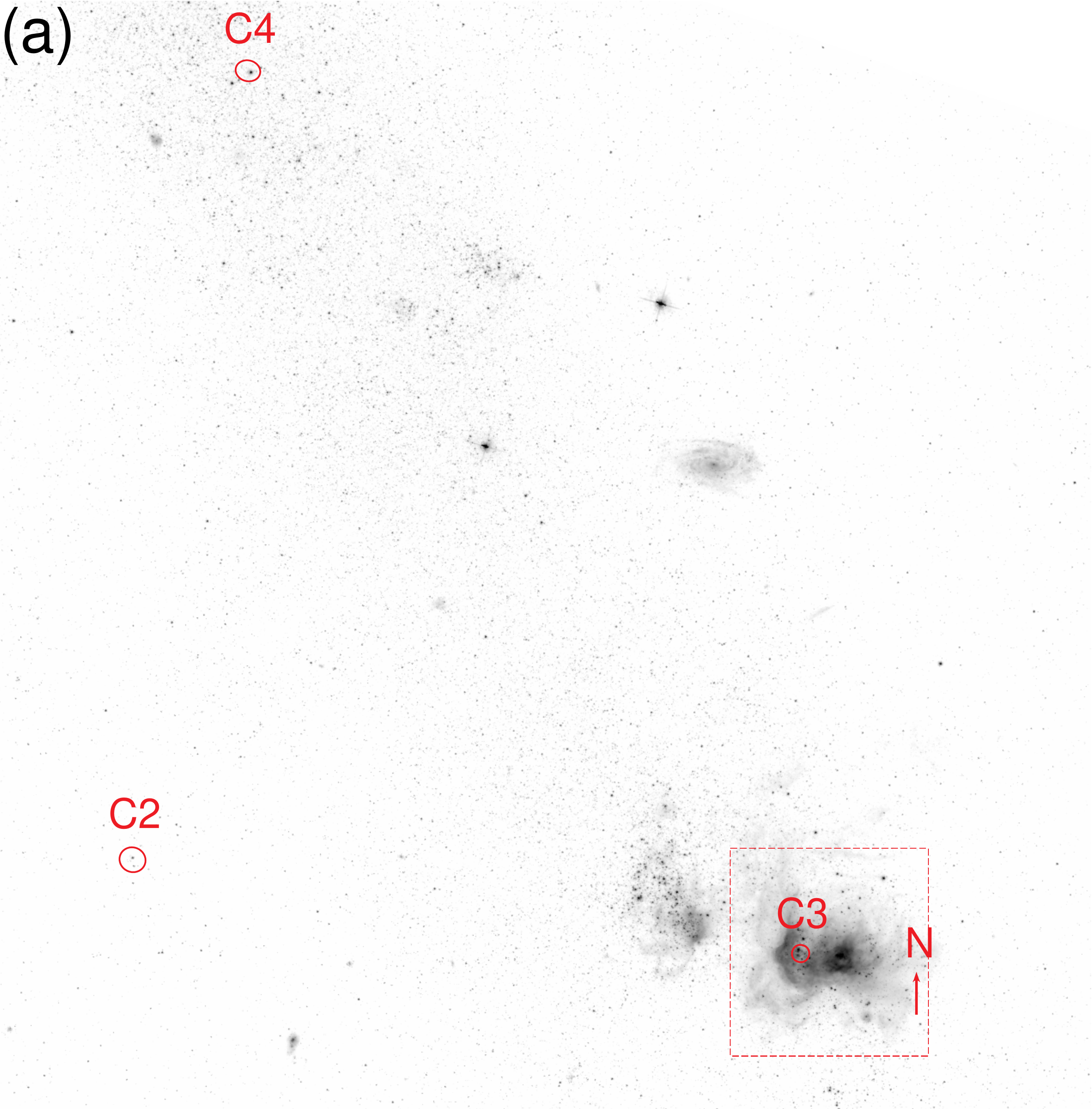}}
\caption{Optical and X-ray images of NGC~2366. The red circle labeled `C3' shows the 1$\arcsec$ radius circle used to extract counts for the point-like Chandra source C3 and is the same in each panel. To aid in orienting between the images, the red dashed box and red arrow pointing North with a length of 5$\arcsec$ are the same in each panel.
Panel \textbf{(a)} - HST/ACS V-band/F555W image of NGC~2366. The positions of Chandra X-ray sources C2 and C4 are shown as red ellipses. The ellipses are those found using {\tt wavdetect}.
\textbf{(b)} - Zoom in of the HST/ACS V-band/F555W image on the region containing the star-forming knots A and B, which are labeled.
\textbf{(c)} - Chandra X-ray image in the region near C3 in the 0.5-7~keV band. The black circle labeled X3 shows the extent of XMMU J072843.4+691123. The skyblue dashed annulus shows the region used to estimate the background in the Chandra image.
} \label{fig:images}
\end{minipage}
\hspace{0.04\textwidth}
\begin{minipage}[c]{0.48\textwidth}
\centerline{\includegraphics[width=\textwidth]{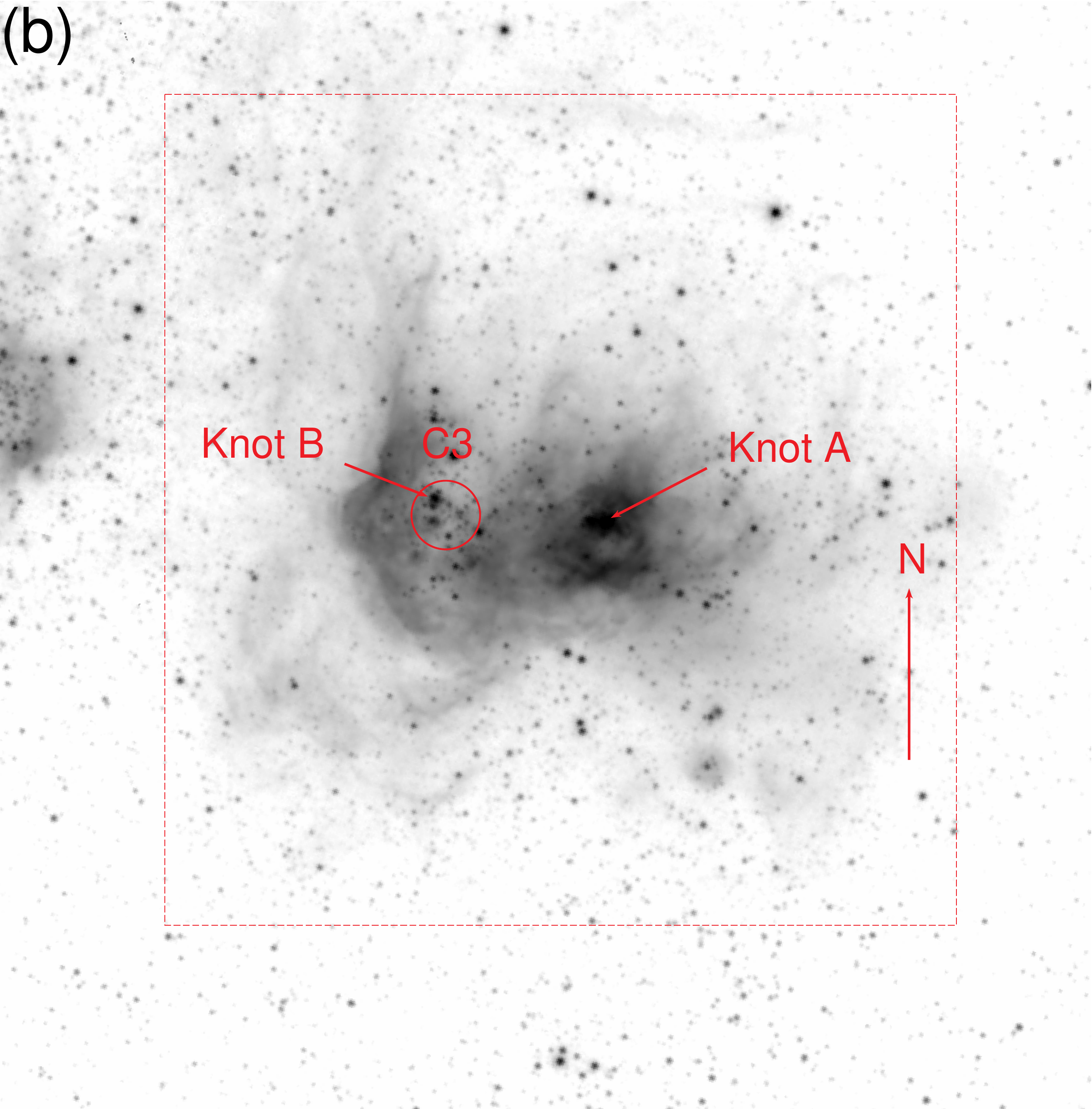}}
\vspace{8pt}
\centerline{\includegraphics[width=0.9\textwidth]{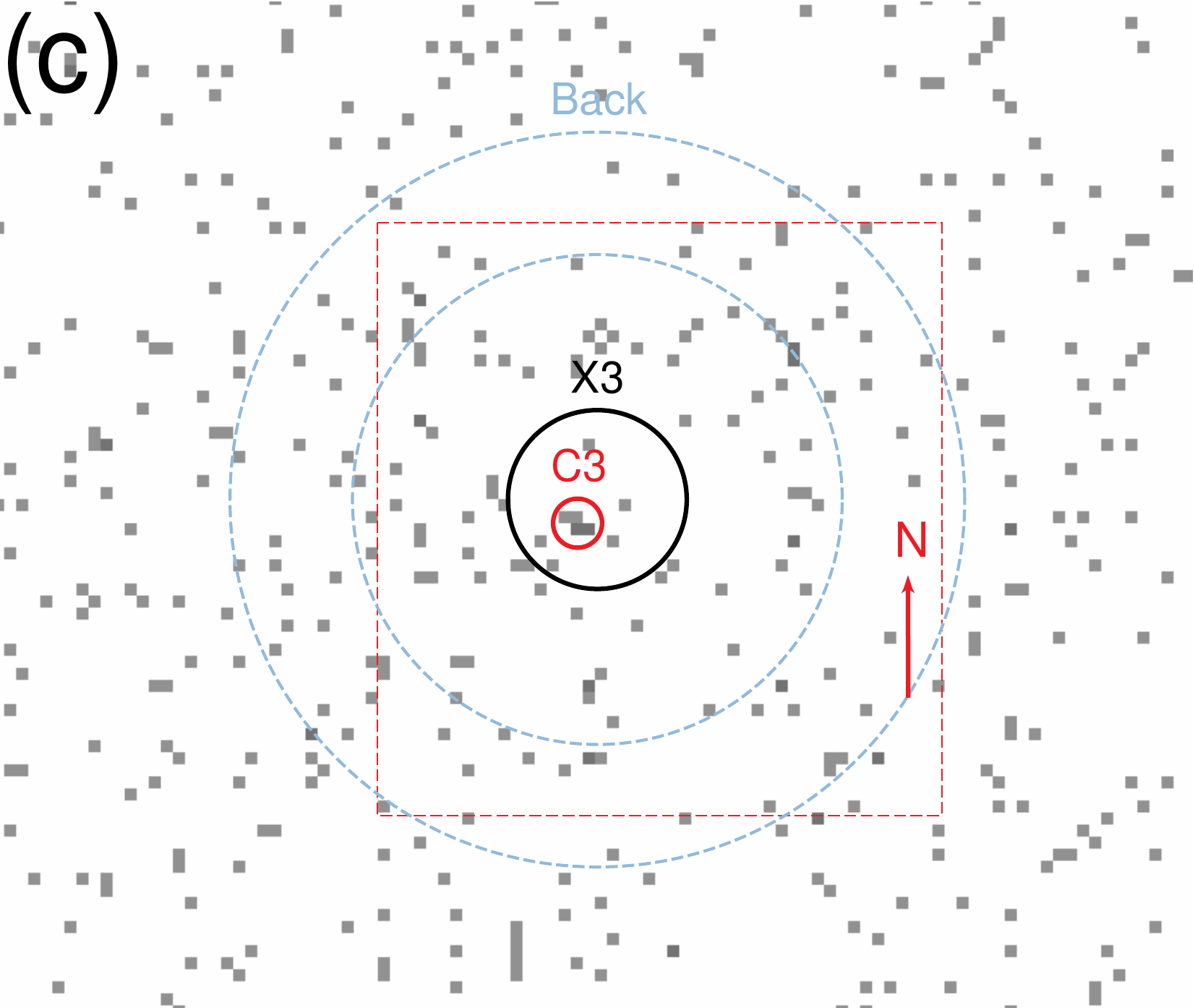}}
\end{minipage}
\vspace{12pt}
\end{figure*}

\section{Observations and Analysis}
\label{sec:obs}

Our approved Chandra exposure time was divided into five observations for operational reasons. The dates and exposures for the individual observations are listed in Table~\ref{tab:obs}. The observations spanned almost two years and have a total exposure of 117.2~ks. We used CIAO version 4.15 \citep{Fruscione2006} and CALDB version 4.10.7 for our analysis. We analyzed data only for the ACIS S3 chip where the target galaxy is located.

We corrected the absolute astrometry of the first observation using the USNO-A2.0 catalog. We used the {\tt wavdetect} tool in CIAO to find X-ray sources and then used {\tt wcs\_match} to find matches with optical sources in the USNO-A2.0 catalog. Four matches were identified. After applying a coordinate translation using {\tt wcs\_update}, the average residual between X-ray and optical positions was 0.34$\arcsec$. We aligned the subsequent observations to the astrometry-corrected first observation. There were at least 9 matched X-ray sources in each pair of observations. 

The Chandra sources detected within NGC~2366 are given in Table~\ref{tab:sources}. Figure~\ref{fig:images}-a shows the Chandra source positions overlaid on a V-band image taken with HST using the Advanced Camera for Surveys (ACS) with the F555W/V-band filter. There are apparent optical counterparts for the sources C2 and C4. These two sources provide a check on the astrometry. The position of each X-ray source lies within 0.33$\arcsec$ of its optical counterpart.

We detect three of the four sources reported from the XMM-Newton observation in 2002 \citep{Thuan2014}. The Chandra source C2 corresponds to the XMM-Newton source XMMU J072858.2+691134 and is likely a background galaxy hosting an active nucleus as noted in \citet{Thuan2014}. C4 corresponds to XMMU J072855.4+691305 and is coincident with a bright and compact H~II region although it could be a foreground star \citep{Thuan2014}. We do not detect a Chandra counterpart to XMMU J072830.4+691132 which is noted to be a very marginal detection \citep{Thuan2014}. The Chandra source C3 is discussed in the next section.

\begin{table}
{\centering
\caption{Chandra X-ray sources in NGC 2366}
\begin{tabular}{clll} \hline
  Source & RA          & DEC        & XMMU Source      \\ \hline
  $-$/X1 & $-$         & $-$        & J072830.4+691132 \\  
  C2/X2  & 07 28 58.14 & 69 11 33.0 & J072858.2+691134 \\
  C3/X3  & 07 28 43.53 & 69 11 22.1 & J072843.4+691123 \\
  C4/X4  & 07 28 55.62 & 69 13 05.0 & J072855.4+691305 \\
\end{tabular} }
\label{tab:sources}
\end{table}

\section{Results}
\label{sec:results}

Using XMM-Newton, \citet{Thuan2014} found an X-ray source located near the star-forming knots in NGC~2366  XMMU J072843.4+691123 referred to here as `X3', with a luminosity of $8 \times 10^{36} \rm \, erg \, s^{-1}$. Our Chandra observation reveals a single X-ray source within the error circle of X3 that we refer to as `C3', see Figure~\ref{fig:images}-c. We used a 1$\arcsec$ radius source region suitable for a point-like source and found 6 counts in the 0.5-7~keV band. We used the annular region shown in the figure to estimate the background level; only 0.62 counts are expected from the background within the source region. Using the Poisson distribution, the chance probability of finding 6 or more counts with 0.62 counts expected is $4.8 \times 10^{-5}$. Allowing for 12 trials, estimated from the ratio of the areas in the XMM and Chandra source regions, the post-trials chance probability is $5.8 \times 10^{-4}$. Thus, the source is confidently detected. The average, background-subtracted source count rate in the 0.5-7~keV band is $4.6 \times 10^{-5} \rm \, counts \, s^{-1}$. We find no significant evidence for variability between the observations, but this is limited by the small number of counts. 

We also examined the Chandra counts within the source region for XMMU~J072843.4+691123. We find 11 counts and the estimated background is 7.7 counts. The chance probability of detecting 11 counts given the background level is 0.15. The flux from the full region is consistent with the flux from C3. Thus, there is no evidence for X-ray emission within the region of X3 beyond that in the point-like source C3.

The effective area of the ACIS-S at the time of our observations drops sharply below 1~keV. Indeed, all of the photons detected from C3 are at energies above 1~keV. To examine the potential variability of the source, we use the XMM flux in the 1-12~keV band as reported in the Fourth XMM-Newton Serendipitous Source Catalog, Thirteenth Data Release \citep{Webb2020} for 4XMM J072843.4+691122 which is identified with X3. The absorbed flux from the 4XMM catalog is $1.5 \times 10^{-15} \rm \, erg \, cm^{-2} \, s^{-1}$ in this band. We use a powerlaw spectrum with a photon index $\Gamma = 2$ and an absorption column depth of $N_{\rm H} = 6.2 \times 10^{20} \rm \, cm^{-2}$, which is the Galactic value towards Mrk~71, to estimate the source flux seen with Chandra. The Chandra flux in the 1-12~keV band is $5.2 \times 10^{-16} \rm \, erg \, cm^{-2} \, s^{-1}$, a factor of $\sim$3 lower than the XMM flux in the same band. To test the statistical significance of the variability, we compare the Chandra counts in the small (1$\arcsec$ radius) region, since a source must be point-like to be variable on time scales of $\sim$10 years at the distance to NGC~2366. If the source flux were constant then 17.3 counts would be expected for Chandra. The chance probability of observing 6 or fewer counts is 0.0017, thus the variability is significant at the 99.8\% confidence level.

The 4XMM source flux in the soft 0.2-1~keV band is $4.3 \times 10^{-15} \rm \, erg \, cm^{-2} \, s^{-1}$. This extended emission should be constant, but is not detected by Chandra due to the cutoff in response of ACIS-S at low energies. Thus, the XMM-Newton and Chandra results combined suggest the presence of a hard, variable, point-like source and soft extended emission.

\section{Discussion}
\label{sec:discussion}

The superb angular resolution of Chandra enables improved localization of the X-ray emission from the star-forming knots in NGC~2366 that clearly associates the emission with knot~B, see Fig.~\ref{fig:images}-b. The Chandra data show only a single point source, C3. The source is variable by a factor of $\sim 3$ with a luminosity in the range of one to a few $10^{36} \rm \, erg \, s^{-1}$, consistent with origin of the emission from an X-ray binary.

The location of C3 is consistent with the apex of the conical outflow of material in NGC~2366 found by \citet{Micheva2019}. This raises the question of whether the X-ray binary may contribute to the kinetic energy needed to power the outflow. The properties of C3 are similar to those of the microquasar, S26, in the galaxy NGC~7793 that has a powerful outflow with a kinetic power of a few  $10^{40} \rm \, erg \, s^{-1}$ \citep{Pakull2010}. S26 has a hard X-ray core identified with the X-ray binary and soft X-ray emission extended over about $15\arcsec$ ($\sim$300~pc) thought to be due to optically thin thermal plasma \citep{Pakull2010}. This is similar to C3/X3 which contains a hard, point source and soft X-ray emission extended over $\sim$150~pc. Both the luminosity and extent of C3/X3 are comparable to S26, but a factor of a few smaller. The location of C3 and the similarity of the properties of C3/X3 to S26 suggests that an X-ray binary could contribute to the mechanical power needed to drive the outflow.

An important question is whether the binary can provide sufficient energy to power the outflow. The kinetic energy in the outflow can be estimated from its speed and mass. \citet{Micheva2019} find a Mach number near 5 at knot B and an average sound speed of $\sim$10~km/s. Thus, the outflow speed is $\sim$50~km/s. They describe the outflow as a cone with an opening angle of $25\arcdeg$. Approximating the cone as having a length of 100~pc and a uniform electron density of $\sim 100 \rm \, cm^{-3}$ (from their Figure~6 and Table~3), the mass of material in the outflow is then $\sim 2 \times 10^{38}$~g. The outflow speed given above then implies an kinetic energy in the outflow of $\sim 3 \times 10^{51}$~erg. We note this is smaller than the kinetic energy estimate in \citet{Roy1992} because we consider only the mass in the conical outflow. The time for material to travel the length of the outflow is 2~Myr. The average kinetic power over this duration is $\sim 5 \times 10^{37} \rm \, erg \, s^{-1}$. This is a factor of a few larger than the X-ray luminosity of the binary. However, the mechanical power output by X-ray binaries can exceed the X-ray luminosity. E.g., the jet power of S26 in NGC~7793 is $10^{4}$ times larger than the X-ray emission from the binary \citep{Pakull2010} while kinetic power of SS~433 exceeds the X-ray luminosity by a factor of 1000 \citep{Marshall2002}.

The stellar population within the cluster provides an alternative means to power the outflow. \citet{Drissen2000} performed stellar population synthesis modeling of the UV spectrum of knot B. From a sequence of models using a Salpeter initial mass function (IMF), a metallicity of $0.1 Z_{\odot}$, and an upper mass cutoff of $80 M_{\odot}$, the best fitting model gives an age of 2.5~Myr and a total kinetic energy returned to the ISM since the beginning of the starburst on the order of $5 \times 10^{50}$~erg.  This is a factor of several smaller than needed to drive the outflow. \citet{Drissen2000} suggest that massive stars and supernova explosions outside knot B may contribute to a broader outflow, but this would be inconsistent with the conical outflow with an apex at knot~B measured by \citet{Micheva2019}. With the young age, the synthesis predicts no Wolf-Rayet (WR) stars in contrast to the observation of several WR stars within knot~B. The presence of Wolf-Rayet stars indicates an age of 3-5~Myr \citep{Micheva2019}. The older ages would allow occurrence of a few supernovae which could be sufficient to power the outflow \citep{Leitherer1999}. A key question is whether any supernovae have occurred. \citet{Chomiuk2009} measured a radio spectral index of $-0.10 \pm 0.12$ from knot~B (source ID N2366-11) that indicates bremsstrahlung emission consistent with an \ion{H}{2} region rather than nonthermal emission as expected from supernova remnants. Also, the emission line ratio [\ion{S}{2}]/H$\alpha$, which is tracer for shock heating in supernova remnants, is low in knot~B \citep{Micheva2019}. These results suggest that shock heating is not energetically important, but do not rule out occurrence of a few supernovae in knot~B.

The first supernovae occuring in knot~B likely had massive progenitors since IMF upper mass limits of at least $60 M_{\odot}$ are needed to reproduce the observed UV spectrum \citep{Drissen2000}. Such massive stars may form black holes directly with a weak or no supernova explosion \citep{Fryer1999,Heger2003}. The occurrence of such failed supernovae is supported by observations of the disappearance of massive stars \citep{Adams2017,Smartt2015}. Thus, it is possible an X-ray binary formed in knot~B without a supernova.

In conclusion, the conical outflow from knot B is thought to create a low density channel in the interstellar medium enabling LyC escape \citep{Micheva2019}. We have detected a point-like X-ray source coincident with knot B. This single X-ray binary could provide sufficient mechanical energy over the lifetime of the outflow to provide the needed kinetic energy. This suggests a potential link between outflows enabling LyC escape and accreting compact objects.

\begin{acknowledgments}

We thank the anonymous referee for thoughtful comments that improved the manuscript.
Support for this work was provided by the National Aeronautics and Space Administration through the Chandra X-ray Observatory Center, which is operated by the Smithsonian Astrophysical Observatory for and on behalf of the National Aeronautics and Space Administration under contract NAS8-03060.
We made use of software provided by the Chandra X-ray Center (CXC) in the CIAO application package and datasets obtained by the Chandra X-ray Observatory contained in the Chandra Data Collection (CDC) \dataset[doi.org/10.25574/cdc.242]{\doi{doi.org/10.25574/cdc.242}}.
We made use of HST data, obtained from the Mikulski Archive for Space Telescopes (MAST) at the Space Telescope Science Institute, contained in \dataset[doi:10.17909/6hfz-rj48]{\doi{10.17909/6hfz-rj48}}.
We made use of data obtained from the 4XMM XMM-Newton serendipitous source catalogue compiled by the ten institutes of the XMM-Newton Survey Science Centre selected by ESA.

\end{acknowledgments}

%
\facilities{Chandra, XMM-Newton, HST}

\software{
\textsc{ds9} \citep{ds9},
\textsc{AstroPy} \citep{astropy2013,astropy2018,astropy2022}
}

\bibliography{ngc2366refs}

\begin{thebibliography}{}
\expandafter\ifx\csname natexlab\endcsname\relax\def\natexlab#1{#1}\fi
\providecommand{\url}[1]{\href{#1}{#1}}
\providecommand{\dodoi}[1]{doi:~\href{http://doi.org/#1}{\nolinkurl{#1}}}
\providecommand{\doeprint}[1]{\href{http://ascl.net/#1}{\nolinkurl{http://ascl.net/#1}}}
\providecommand{\doarXiv}[1]{\href{https://arxiv.org/abs/#1}{\nolinkurl{https://arxiv.org/abs/#1}}}

\bibitem[{{Adams} {et~al.}(2017){Adams}, {Kochanek}, {Gerke}, {Stanek}, \&
  {Dai}}]{Adams2017}
{Adams}, S.~M., {Kochanek}, C.~S., {Gerke}, J.~R., {Stanek}, K.~Z., \& {Dai},
  X. 2017, \mnras, 468, 4968, \dodoi{10.1093/mnras/stx816}

\bibitem[{{Astropy Collaboration} {et~al.}(2013){Astropy Collaboration},
  {Robitaille}, {Tollerud}, {Greenfield}, {Droettboom}, {Bray}, {Aldcroft},
  {Davis}, {Ginsburg}, {Price-Whelan}, {Kerzendorf}, {Conley}, {Crighton},
  {Barbary}, {Muna}, {Ferguson}, {Grollier}, {Parikh}, {Nair}, {Unther},
  {Deil}, {Woillez}, {Conseil}, {Kramer}, {Turner}, {Singer}, {Fox}, {Weaver},
  {Zabalza}, {Edwards}, {Azalee Bostroem}, {Burke}, {Casey}, {Crawford},
  {Dencheva}, {Ely}, {Jenness}, {Labrie}, {Lim}, {Pierfederici}, {Pontzen},
  {Ptak}, {Refsdal}, {Servillat}, \& {Streicher}}]{astropy2013}
{Astropy Collaboration}, {Robitaille}, T.~P., {Tollerud}, E.~J., {et~al.} 2013,
  \aap, 558, A33, \dodoi{10.1051/0004-6361/201322068}

\bibitem[{{Astropy Collaboration} {et~al.}(2018){Astropy Collaboration},
  {Price-Whelan}, {Sip{\H{o}}cz}, {G{\"u}nther}, {Lim}, {Crawford}, {Conseil},
  {Shupe}, {Craig}, {Dencheva}, {Ginsburg}, {VanderPlas}, {Bradley},
  {P{\'e}rez-Su{\'a}rez}, {de Val-Borro}, {Aldcroft}, {Cruz}, {Robitaille},
  {Tollerud}, {Ardelean}, {Babej}, {Bach}, {Bachetti}, {Bakanov}, {Bamford},
  {Barentsen}, {Barmby}, {Baumbach}, {Berry}, {Biscani}, {Boquien}, {Bostroem},
  {Bouma}, {Brammer}, {Bray}, {Breytenbach}, {Buddelmeijer}, {Burke},
  {Calderone}, {Cano Rodr{\'\i}guez}, {Cara}, {Cardoso}, {Cheedella}, {Copin},
  {Corrales}, {Crichton}, {D'Avella}, {Deil}, {Depagne}, {Dietrich}, {Donath},
  {Droettboom}, {Earl}, {Erben}, {Fabbro}, {Ferreira}, {Finethy}, {Fox},
  {Garrison}, {Gibbons}, {Goldstein}, {Gommers}, {Greco}, {Greenfield},
  {Groener}, {Grollier}, {Hagen}, {Hirst}, {Homeier}, {Horton}, {Hosseinzadeh},
  {Hu}, {Hunkeler}, {Ivezi{\'c}}, {Jain}, {Jenness}, {Kanarek}, {Kendrew},
  {Kern}, {Kerzendorf}, {Khvalko}, {King}, {Kirkby}, {Kulkarni}, {Kumar},
  {Lee}, {Lenz}, {Littlefair}, {Ma}, {Macleod}, {Mastropietro}, {McCully},
  {Montagnac}, {Morris}, {Mueller}, {Mumford}, {Muna}, {Murphy}, {Nelson},
  {Nguyen}, {Ninan}, {N{\"o}the}, {Ogaz}, {Oh}, {Parejko}, {Parley}, {Pascual},
  {Patil}, {Patil}, {Plunkett}, {Prochaska}, {Rastogi}, {Reddy Janga},
  {Sabater}, {Sakurikar}, {Seifert}, {Sherbert}, {Sherwood-Taylor}, {Shih},
  {Sick}, {Silbiger}, {Singanamalla}, {Singer}, {Sladen}, {Sooley},
  {Sornarajah}, {Streicher}, {Teuben}, {Thomas}, {Tremblay}, {Turner},
  {Terr{\'o}n}, {van Kerkwijk}, {de la Vega}, {Watkins}, {Weaver}, {Whitmore},
  {Woillez}, {Zabalza}, \& {Astropy Contributors}}]{astropy2018}
{Astropy Collaboration}, {Price-Whelan}, A.~M., {Sip{\H{o}}cz}, B.~M., {et~al.}
  2018, \aj, 156, 123, \dodoi{10.3847/1538-3881/aabc4f}

\bibitem[{{Astropy Collaboration} {et~al.}(2022){Astropy Collaboration},
  {Price-Whelan}, {Lim}, {Earl}, {Starkman}, {Bradley}, {Shupe}, {Patil},
  {Corrales}, {Brasseur}, {N{\"o}the}, {Donath}, {Tollerud}, {Morris},
  {Ginsburg}, {Vaher}, {Weaver}, {Tocknell}, {Jamieson}, {van Kerkwijk},
  {Robitaille}, {Merry}, {Bachetti}, {G{\"u}nther}, {Aldcroft},
  {Alvarado-Montes}, {Archibald}, {B{\'o}di}, {Bapat}, {Barentsen},
  {Baz{\'a}n}, {Biswas}, {Boquien}, {Burke}, {Cara}, {Cara}, {Conroy},
  {Conseil}, {Craig}, {Cross}, {Cruz}, {D'Eugenio}, {Dencheva}, {Devillepoix},
  {Dietrich}, {Eigenbrot}, {Erben}, {Ferreira}, {Foreman-Mackey}, {Fox},
  {Freij}, {Garg}, {Geda}, {Glattly}, {Gondhalekar}, {Gordon}, {Grant},
  {Greenfield}, {Groener}, {Guest}, {Gurovich}, {Handberg}, {Hart},
  {Hatfield-Dodds}, {Homeier}, {Hosseinzadeh}, {Jenness}, {Jones}, {Joseph},
  {Kalmbach}, {Karamehmetoglu}, {Ka{\l}uszy{\'n}ski}, {Kelley}, {Kern},
  {Kerzendorf}, {Koch}, {Kulumani}, {Lee}, {Ly}, {Ma}, {MacBride}, {Maljaars},
  {Muna}, {Murphy}, {Norman}, {O'Steen}, {Oman}, {Pacifici}, {Pascual},
  {Pascual-Granado}, {Patil}, {Perren}, {Pickering}, {Rastogi}, {Roulston},
  {Ryan}, {Rykoff}, {Sabater}, {Sakurikar}, {Salgado}, {Sanghi}, {Saunders},
  {Savchenko}, {Schwardt}, {Seifert-Eckert}, {Shih}, {Jain}, {Shukla}, {Sick},
  {Simpson}, {Singanamalla}, {Singer}, {Singhal}, {Sinha}, {Sip{\H{o}}cz},
  {Spitler}, {Stansby}, {Streicher}, {{\v{S}}umak}, {Swinbank}, {Taranu},
  {Tewary}, {Tremblay}, {de Val-Borro}, {Van Kooten}, {Vasovi{\'c}}, {Verma},
  {de Miranda Cardoso}, {Williams}, {Wilson}, {Winkel}, {Wood-Vasey}, {Xue},
  {Yoachim}, {Zhang}, {Zonca}, \& {Astropy Project Contributors}}]{astropy2022}
{Astropy Collaboration}, {Price-Whelan}, A.~M., {Lim}, P.~L., {et~al.} 2022,
  \apj, 935, 167, \dodoi{10.3847/1538-4357/ac7c74}

\bibitem[{{Bluem} {et~al.}(2019){Bluem}, {Kaaret}, {Prestwich}, \&
  {Brorby}}]{Bluem2019}
{Bluem}, J., {Kaaret}, P., {Prestwich}, A., \& {Brorby}, M. 2019, \mnras, 487,
  4093, \dodoi{10.1093/mnras/stz1574}

\bibitem[{{Cardamone} {et~al.}(2009){Cardamone}, {Schawinski}, {Sarzi},
  {Bamford}, {Bennert}, {Urry}, {Lintott}, {Keel}, {Parejko}, {Nichol},
  {Thomas}, {Andreescu}, {Murray}, {Raddick}, {Slosar}, {Szalay}, \&
  {Vandenberg}}]{Cardamone2009}
{Cardamone}, C., {Schawinski}, K., {Sarzi}, M., {et~al.} 2009, \mnras, 399,
  1191, \dodoi{10.1111/j.1365-2966.2009.15383.x}

\bibitem[{{Chomiuk} \& {Wilcots}(2009)}]{Chomiuk2009}
{Chomiuk}, L., \& {Wilcots}, E.~M. 2009, \aj, 137, 3869,
  \dodoi{10.1088/0004-6256/137/4/3869}

\bibitem[{{Drissen} {et~al.}(2000){Drissen}, {Roy}, {Robert}, {Devost}, \&
  {Doyon}}]{Drissen2000}
{Drissen}, L., {Roy}, J.-R., {Robert}, C., {Devost}, D., \& {Doyon}, R. 2000,
  \aj, 119, 688, \dodoi{10.1086/301204}

\bibitem[{{Fruscione} {et~al.}(2006){Fruscione}, {McDowell}, {Allen},
  {Brickhouse}, {Burke}, {Davis}, {Durham}, {Elvis}, {Galle}, {Harris},
  {Huenemoerder}, {Houck}, {Ishibashi}, {Karovska}, {Nicastro}, {Noble},
  {Nowak}, {Primini}, {Siemiginowska}, {Smith}, \& {Wise}}]{Fruscione2006}
{Fruscione}, A., {McDowell}, J.~C., {Allen}, G.~E., {et~al.} 2006, in Society
  of Photo-Optical Instrumentation Engineers (SPIE) Conference Series, Vol.
  6270, Observatory Operations: Strategies, Processes, and Systems, ed. D.~R.
  {Silva} \& R.~E. {Doxsey}, 62701V, \dodoi{10.1117/12.671760}

\bibitem[{{Fryer}(1999)}]{Fryer1999}
{Fryer}, C.~L. 1999, \apj, 522, 413, \dodoi{10.1086/307647}

\bibitem[{{Gallo} {et~al.}(2005){Gallo}, {Fender}, {Kaiser}, {Russell},
  {Morganti}, {Oosterloo}, \& {Heinz}}]{Gallo2005}
{Gallo}, E., {Fender}, R., {Kaiser}, C., {et~al.} 2005, \nat, 436, 819,
  \dodoi{10.1038/nature03879}

\bibitem[{{Gross} {et~al.}(2021){Gross}, {Prestwich}, \& {Kaaret}}]{Gross2021}
{Gross}, A.~C., {Prestwich}, A., \& {Kaaret}, P. 2021, \mnras, 505, 610,
  \dodoi{10.1093/mnras/stab1331}

\bibitem[{{Hayes} {et~al.}(2010){Hayes}, {{\"O}stlin}, {Schaerer}, {Mas-Hesse},
  {Leitherer}, {Atek}, {Kunth}, {Verhamme}, {de Barros}, \&
  {Melinder}}]{Hayes2010}
{Hayes}, M., {{\"O}stlin}, G., {Schaerer}, D., {et~al.} 2010, \nat, 464, 562,
  \dodoi{10.1038/nature08881}

\bibitem[{{Heger} {et~al.}(2003){Heger}, {Fryer}, {Woosley}, {Langer}, \&
  {Hartmann}}]{Heger2003}
{Heger}, A., {Fryer}, C.~L., {Woosley}, S.~E., {Langer}, N., \& {Hartmann},
  D.~H. 2003, \apj, 591, 288, \dodoi{10.1086/375341}

\bibitem[{{Izotov} {et~al.}(2011){Izotov}, {Guseva}, \&
  {Thuan}}]{Izotov2011ApJ728}
{Izotov}, Y.~I., {Guseva}, N.~G., \& {Thuan}, T.~X. 2011, \apj, 728, 161,
  \dodoi{10.1088/0004-637X/728/2/161}

\bibitem[{{Kaaret} {et~al.}(2022){Kaaret}, {Bluem}, \&
  {Prestwich}}]{Kaaret2022}
{Kaaret}, P., {Bluem}, J., \& {Prestwich}, A.~H. 2022, \mnras, 511, L8,
  \dodoi{10.1093/mnrasl/slab127}

\bibitem[{{Kaaret} {et~al.}(2017){Kaaret}, {Brorby}, {Casella}, \&
  {Prestwich}}]{Kaaret2017}
{Kaaret}, P., {Brorby}, M., {Casella}, L., \& {Prestwich}, A.~H. 2017, \mnras,
  471, 4234, \dodoi{10.1093/mnras/stx1945}

\bibitem[{{Kimm} \& {Cen}(2014)}]{Kimm2014}
{Kimm}, T., \& {Cen}, R. 2014, \apj, 788, 121,
  \dodoi{10.1088/0004-637X/788/2/121}

\bibitem[{{Leitherer} {et~al.}(1999){Leitherer}, {Schaerer}, {Goldader},
  {Delgado}, {Robert}, {Kune}, {de Mello}, {Devost}, \&
  {Heckman}}]{Leitherer1999}
{Leitherer}, C., {Schaerer}, D., {Goldader}, J.~D., {et~al.} 1999, \apjs, 123,
  3, \dodoi{10.1086/313233}

\bibitem[{{Marshall} {et~al.}(2002){Marshall}, {Canizares}, \&
  {Schulz}}]{Marshall2002}
{Marshall}, H.~L., {Canizares}, C.~R., \& {Schulz}, N.~S. 2002, \apj, 564, 941,
  \dodoi{10.1086/324398}

\bibitem[{{Micheva} {et~al.}(2019){Micheva}, {Christian Herenz}, {Roth},
  {{\"O}stlin}, \& {Girichidis}}]{Micheva2019}
{Micheva}, G., {Christian Herenz}, E., {Roth}, M.~M., {{\"O}stlin}, G., \&
  {Girichidis}, P. 2019, \aap, 623, A145, \dodoi{10.1051/0004-6361/201834838}

\bibitem[{{Pakull} {et~al.}(2010){Pakull}, {Soria}, \& {Motch}}]{Pakull2010}
{Pakull}, M.~W., {Soria}, R., \& {Motch}, C. 2010, \nat, 466, 209,
  \dodoi{10.1038/nature09168}

\bibitem[{{Prestwich} {et~al.}(2015){Prestwich}, {Jackson}, {Kaaret}, {Brorby},
  {Roberts}, {Saar}, \& {Yukita}}]{Prestwich2015}
{Prestwich}, A.~H., {Jackson}, F., {Kaaret}, P., {et~al.} 2015, \apj, 812, 166,
  \dodoi{10.1088/0004-637X/812/2/166}

\bibitem[{{Roy} {et~al.}(1992){Roy}, {Aube}, {McCall}, \& {Dufour}}]{Roy1992}
{Roy}, J.-R., {Aube}, M., {McCall}, M.~L., \& {Dufour}, R.~J. 1992, \apj, 386,
  498, \dodoi{10.1086/171035}

\bibitem[{{Smartt}(2015)}]{Smartt2015}
{Smartt}, S.~J. 2015, \pasa, 32, e016, \dodoi{10.1017/pasa.2015.17}

\bibitem[{{Smithsonian Astrophysical Observatory}(2000)}]{ds9}
{Smithsonian Astrophysical Observatory}. 2000, {SAOImage DS9: A utility for
  displaying astronomical images in the X11 window environment}, Astrophysics
  Source Code Library, record ascl:0003.002.
\newblock \doeprint{0003.002}

\bibitem[{{Thuan} {et~al.}(2014){Thuan}, {Bauer}, \& {Izotov}}]{Thuan2014}
{Thuan}, T.~X., {Bauer}, F.~E., \& {Izotov}, Y.~I. 2014, \mnras, 441, 1841,
  \dodoi{10.1093/mnras/stu716}

\bibitem[{{Tolstoy} {et~al.}(1995){Tolstoy}, {Saha}, {Hoessel}, \&
  {McQuade}}]{Tolstoy1995}
{Tolstoy}, E., {Saha}, A., {Hoessel}, J.~G., \& {McQuade}, K. 1995, \aj, 110,
  1640, \dodoi{10.1086/117637}

\bibitem[{{Webb} {et~al.}(2020){Webb}, {Coriat}, {Traulsen}, {Ballet}, {Motch},
  {Carrera}, {Koliopanos}, {Authier}, {de la Calle}, {Ceballos}, {Colomo},
  {Chuard}, {Freyberg}, {Garcia}, {Kolehmainen}, {Lamer}, {Lin}, {Maggi},
  {Michel}, {Page}, {Page}, {Perea-Calderon}, {Pineau}, {Rodriguez}, {Rosen},
  {Santos Lleo}, {Saxton}, {Schwope}, {Tom{\'a}s}, {Watson}, \&
  {Zakardjian}}]{Webb2020}
{Webb}, N.~A., {Coriat}, M., {Traulsen}, I., {et~al.} 2020, \aap, 641, A136,
  \dodoi{10.1051/0004-6361/201937353}

\end{thebibliography}
\bibliographystyle{aasjournal}

\end{document}